\documentclass[fleqn,10pt]{SelfArx} 

\definecolor{color1}{RGB}{0,0,90} 
\definecolor{color2}{RGB}{0,20,20} 

\usepackage{hyperref} 
\hypersetup{hidelinks,colorlinks,breaklinks=true,urlcolor=color2,citecolor=color1,linkcolor=color1,bookmarksopen=false,pdftitle={Title},pdfauthor={Author},urlcolor=blue}

\usepackage{graphicx}
\usepackage[square]{natbib}
\usepackage{amsmath}
\usepackage{multirow}
\usepackage{subfigure}

\newcommand{\n}[1]{\mathrm{#1}}

\JournalInfo{Published in Scripta Materialia, Vol. 67 (1), 81-84, 2012} 
\Archive{\href{http://dx.doi.org/10.1016/j.scriptamat.2012.03.024}{DOI: 10.1016/j.scriptamat.2012.03.024}} 

\PaperTitle{The sintering behavior of close packed spheres} 

\Authors{R. Bj\o{}rk$^1$, V. Tikare$^2$, H. L. Frandsen$^1$ and N. Pryds$^1$} 
\affiliation{\textsuperscript{1}\textit{Department of Energy Conversion and Storage, Technical University of Denmark - DTU, Frederiksborgvej 399, DK-4000 Roskilde, Denmark}} 
\affiliation{\textsuperscript{2}\textit{Sandia National Laboratory, Albuquerque, NM 87185, New Mexico, USA}} 
\affiliation{*\textbf{Corresponding author}: rabj@dtu.dk} 

\newcommand{\keywordname}{Keywords} 

\Abstract{The sintering behavior of close packed spheres is investigated using a numerical model. The investigated systems are the body centered cubic (BCC), face centered cubic (FCC) and hexagonal closed packed spheres (HCP). The sintering behavior is found to be ideal with no grain growth until full density is reached for all systems. During sintering the grains change shape from spherical to tetrakaidecahedron, similar to the geometry analyzed by Coble (R. L. Coble, J. Appl. Phys. 32 (1961) 787).}

\begin{document}

\flushbottom 

\maketitle 


\thispagestyle{empty} 

The sintering behavior and microstructural evolution of a powder compact is influenced strongly by initial properties, such as the relative density, the particle and pore size distribution and the powder packing. While the influence of the former parameters on the microstructural evolution has been investigated in some detail, the impact of the initial packing of the powder has been mostly overlooked. However, research has shown that the sintering behavior of a powder can be significantly improved, if the powder is regularly packed. This has been shown for monodisperse spherical TiO$_2$ particles \cite{Barringer_1988}, which sintered ten times faster and exhibited almost no grain growth compared to ordinary TiO$_2$. Similar observations has been made for homogeneously packed Al$_2$O$_3$ \cite{Ikegami_1984}, SiO$_2$ \cite{Sacks_1984b}, as well as a number of other materials \cite{Matijevic_1996}. Monodisperse spherical TiO$_2$ particles has been shown to order in face centered cubic (FCC) arrays, while the SiO$_2$ powder orders to form stacked planes of hexagonal close packed particles. Close-packing of monodispersed silica has also been observed \cite{Milne_1993}. Sintering of two dimensional close packing cylinders has also been demonstrated experimentally \cite{Alexander_1957,Liniger_1987,Liniger_1988} and numerically modeled \cite{Braginsky_2005,Tikare_2005} and the sintering of particle clusters in three dimensions have also been studied \cite{Wakai_2006}.

Here we present a detailed analysis of the microstructural evolution of three close packed compacts of spherical mono-size particles. These are the body centered cubic (BCC), face centered cubic (FCC) and hexagonal closed packed (HCP) structures. The packing structures are characterized by their relative density, $\rho$, and the coordination number, $C_\mathrm{n}$, of the grains and pores. The BCC packing initially has $\rho=69.02$\%, and a grain coordination number of eight. The porosity is percolating, but the repeating pore units have a coordination number of six, making the pore unit shape octahedral. The HCP and FCC packings represent the densest packing of spheres possible, and both have a $\rho=74.05$\%. Both have a grain coordination number of twelve, and have the same type of pore structure. This consists of two different kind of repeating pore units, which have a coordination number of either four or six, with the number ratio of 2:1 and the volume ratio between a single four- and six-coordinated pore being 1:6.4 \cite{Hirata_1990}. These pore units are tetrahedral and octahedral in shape respectively.

The sintering behavior of these compacts are analyzed using a three dimensional kinetic Monte Carlo (kMC) sintering model \cite{Tikare_2010, Cardona_2011}. In this model grains and pores are defined on a three dimensional voxel grid and sintering is modeled by minimizing the neighbor interaction energy, $E$, defined such that only unlike neighbors contribute to the system energy. The used numerical model simulates grain growth, pore migration, vacancy formation and diffusion processes, and has previously been used to model the sintering of copper spheres as observed using X-ray tomography \cite{Tikare_2010}. The model has also replicated experimentally observed sintering trends for samples with different initial particle size distributions \cite{Bjoerk_2012a}. Units in the model are arbitrary, and similarly to Bj\o{}rk \emph{et. al.} \cite{Bjoerk_2012a} the simulation temperatures for grain growth and pore migration in the model are chosen as $k_\n{B} T= 1$, while for vacancy formation $k_\n{B} T= 15$. The attempt frequencies were chosen in the ratio 1:1:5 for grain growth, pore migration and vacancy formation, respectively. The values have been chosen such that the modeled samples displayed realistic sintering behavior.

Each of the simulated microstructures consisted of $6\times{}6\times{}6$ particles, which is sufficient to make the center of the sample homogeneous, as the microstructure is completely uniform. The microstructural evolutions of the samples during sintering have been computed during isothermal sintering, and the microstructures have been analyzed at 251 equal intervals in time. The microstructural evolution was only analyzed in the central part of the sample, thus edge effects due to the finite size of the compact can be ignored.

All compacts sintered to full density without grain growth and the sintering behavior of the BCC, FCC and HCP samples are thus ideal. A compact of random close packed (RCP) mono-sized spheres has been shown to sinter to a density of 93\%, at the same temperature, before grain growth occurred \cite{Bjoerk_2012a}. This is similar to experimentally observed values \cite{German_2010}. Controlling the grain growth during sintering is of the utmost importance as most properties of ceramics are enhanced by smaller grain size \cite{Rahaman_2008}. The only difference between the random close packed sample of mono-sized spheres and the systems studied here is the packing. Therefore the ability to sinter without grain growth is caused by regular packing of the powder.

\begin{figure}[!t]
  \centering
  \includegraphics[width=1\columnwidth]{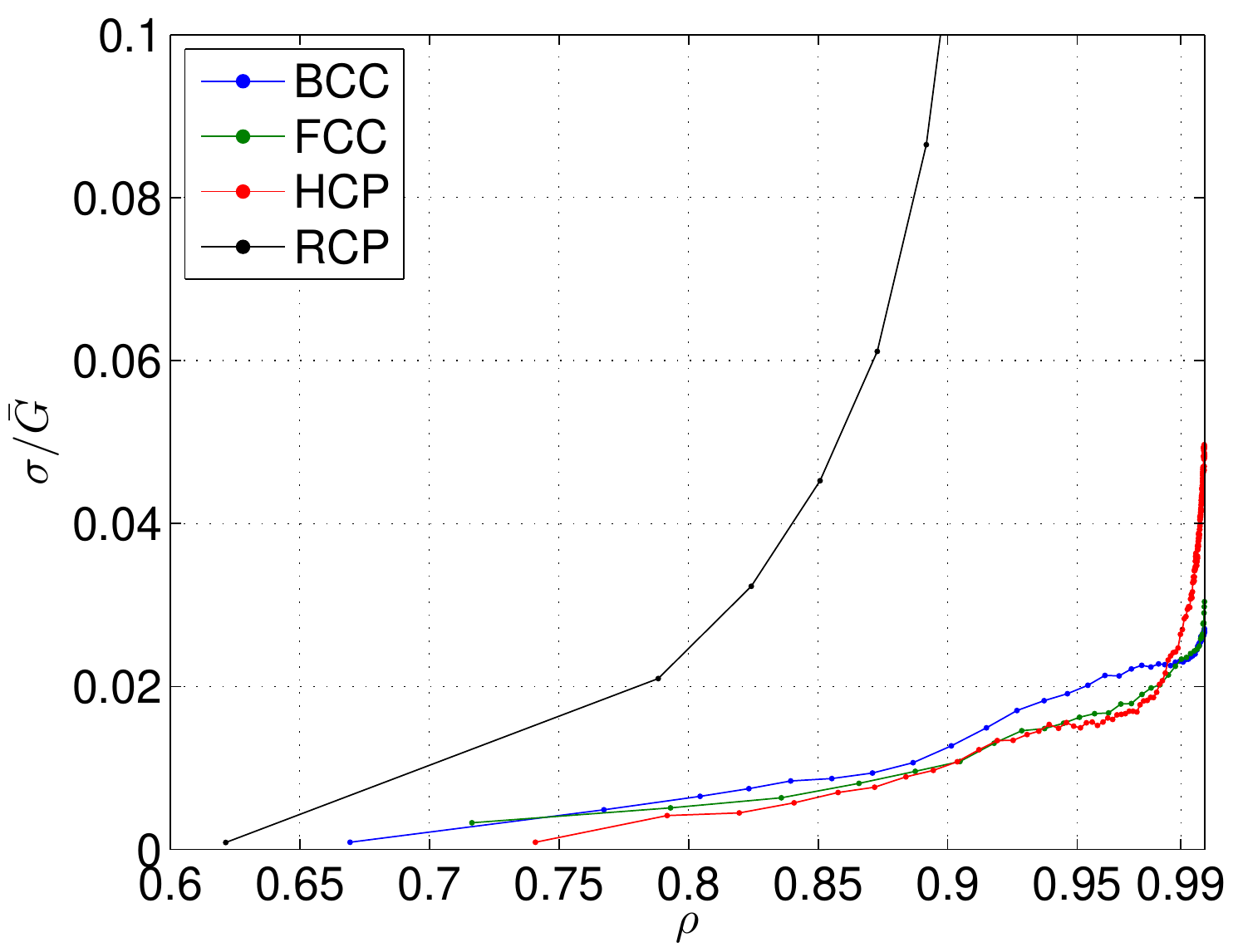}
    \caption{The normalized standard deviation of the grain size, $\sigma/\bar{G}$, for the different packings. The data for the mono-sized random close packed spheres (RCP) are taken from Bj\o{}rk \emph{et. al.} \cite{Bjoerk_2012a}.}
    \label{Fig_Std_dens_regular}
\end{figure}

The standard deviation of the grain size, $\sigma$, normalized by the average grain size is shown in Fig. \ref{Fig_Std_dens_regular}. Here it can clearly be seen that the grain size distributions remains extremely narrow, further showing the ideal sintering behavior of these systems. That the uniformity of the grain size distribution is not caused by the initially uniform particle size can be seen by comparing the evolution of the systems studied here to the grain growth of random close packed mono-sized spheres. At a density of 0.9 the RCP has a normalized standard deviation which is a factor of 10 higher than the BCC, FCC and HCP systems. At a density of 0.95 the factor has risen to 20.

\begin{figure}[!t]
  \centering
  \includegraphics[width=1\columnwidth]{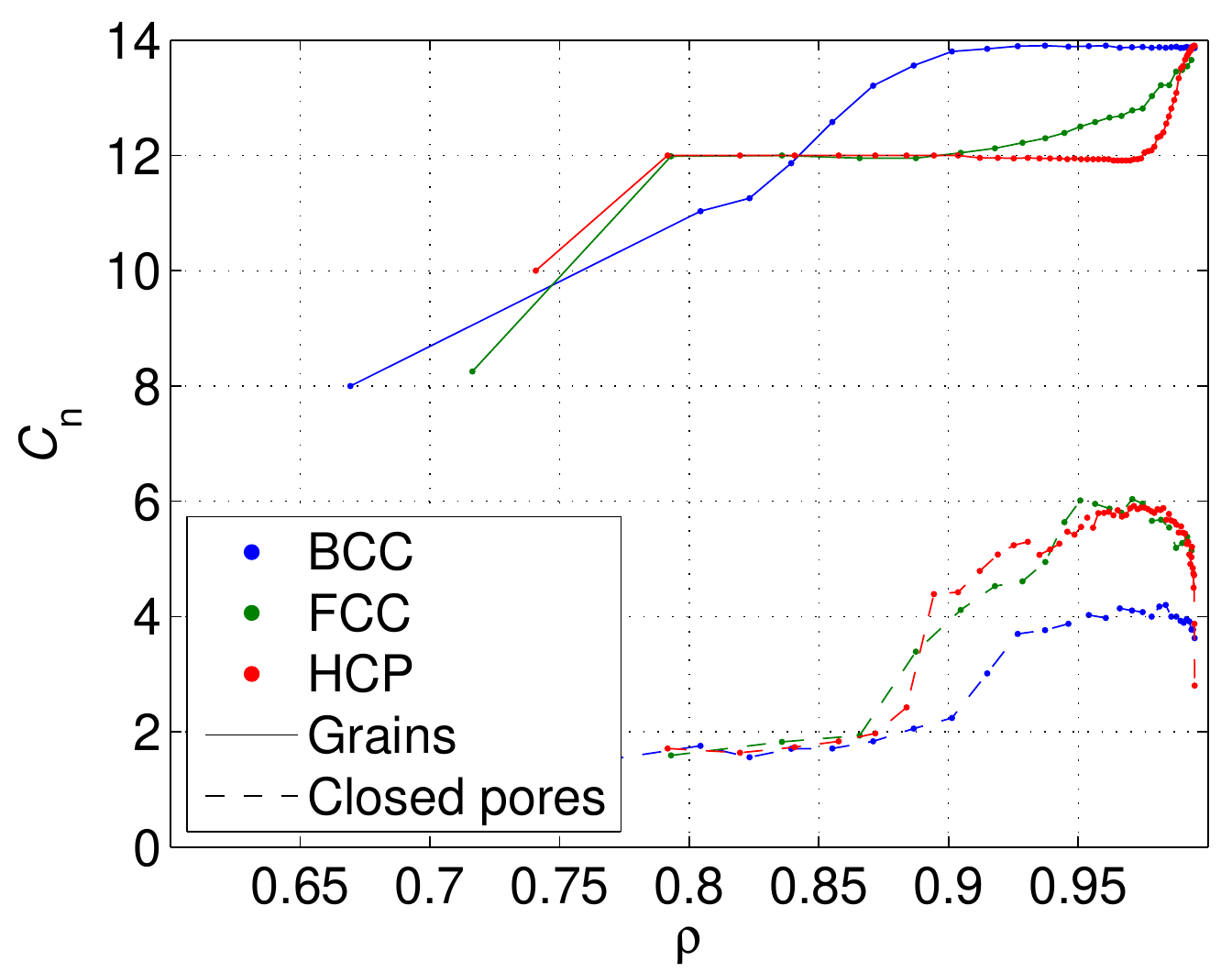}
    \caption{The average coordination number of grains and closed pores as function of relative density for the different packings.}
    \label{Fig_mean_coordination_number}
\end{figure}

The evolution of the microstructure towards full density can be examined by considering the change in coordination number of the grains and closed pores during sintering, as this provides information on the grain and pore shape. Fig. \ref{Fig_mean_coordination_number} shows the average coordination number of grain and closed pores as a function of density.

For the BCC packing, initially all grains have a coordination number of eight, as expected. As sintering progresses the microstructure evolves to an intermediate state, where grains with several different coordination numbers exist. At $\rho=0.9$ almost all grains have evolved to have a coordination number of 14. This is because the porosity gaps between the central sphere in the BCC packing and those directly adjacent in the horizontal and vertical direction have been closed.  The central grain retains contact with the eight diagonal neighbors and come in contact with the six face neighbors in the horizontal and vertical directions; thus, increasing the coordination number to 14. The pores in the BCC system were expected to have a coordination number of six, yet this is not the case. The reason for this behavior is that when the porosity closes at high relative density, the grains along the horizontal and vertical directions have sintered together, reducing the original pore with a coordination number of six into a smaller pore with a coordination number of four.

The grains in the FCC packing is seen to initially have an average coordination numbers of eight, but all grains very quickly evolve to have a coordination number of 12. The reason for the initially lower coordination number is the packing on a finite voxel grid. As sintering progresses the coordination number increases to 14, by closing the porosity in both the top or bottom layer compared to the central sphere. The closed pores have a coordination number of six. Initially many closed pores with a coordination number of four are present, but their aggregate volume is not significant. This is expected as the volume of a $C_\mathrm{n}=4$ pore initially is only 15.6\% of that of a $C_\mathrm{n}=6$ pore in the FCC microstructure.

For the HCP packing initially all grains have a coordination number of ten, which is again caused by the finite voxel resolution. However, all grains very quickly evolve to a coordination number of 12. Then the same pores as those in FCC packing are closed, which increases the coordination number to 14. Unlike FCC the HCP packing transformation of grains from a coordination number of 12 to 14 happens over a very small change in relative density. For the HCP sample, pores with coordination number six dominate. As with the FCC sample closed pores with coordination number four are present initially, but have a low volume and quickly disappear.

In the final stage of sintering all pores are closed. A linear decrease in closed pore volume at high relative density is observed for all samples. This is also observed experimentally for a uniform packed powder, albeit over a larger interval in relative density \cite{Uchida_2001}.

\begin{figure}[!t]
  \centering
  \includegraphics[width=1\columnwidth]{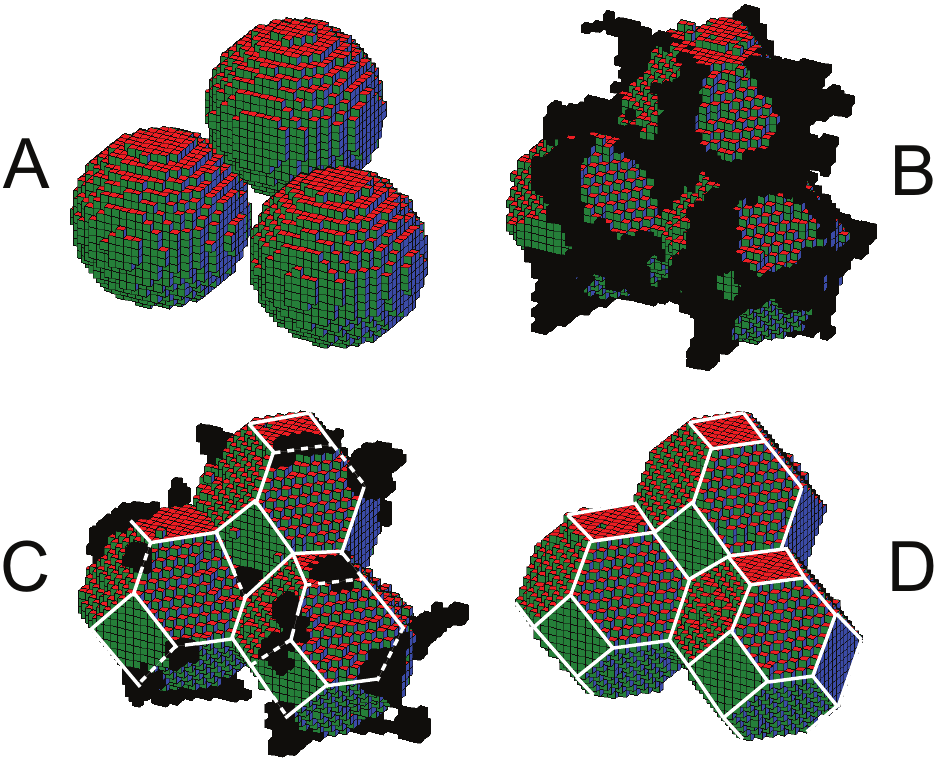}
    \caption{[Color online] The evolution of grains and porosity at times $t=0,6,11,29$ for the BCC structure. The grains are shown as voxels with each face having a different color. Porosity is shown as black voxels. Lines acting as guides to the eye indicate the grains faces. The porosity is not shown for $t=0$, while for $t=6,11$ only selected porosities are shown.}
    \label{Fig_BCC_spheres}
\end{figure}

As mentioned the grains evolve to a coordination number of 14, corresponding to a tetrakaidecahedron shape. The exact shape has been verified to be a truncated octahedron, also referred to as Kelvin's tetrakaidecahedron. The shape of the grains and porosity for the BCC packing are illustrated in Fig. \ref{Fig_BCC_spheres}. The grains are initially spherical, but quickly develop regular faces and the tetrakaidecahedron shape. The final grain structure is similar for the FCC and HCP systems.

The sintering of a system consisting of tetrakaidecahedral particles has been described analytically by Coble \cite{Coble_1961a,Coble_1961b}, whose model has been used to describe the sintering behavior of a substantial number of systems. However, to the authors' knowledge no system has been shown to sinter to the exact morphology assumed by the Coble model.

The sintering behavior of the systems considered here cannot directly be compared to the analytical model derived by Coble, as the porosity is not uniformly distributed along all grain edges, nor is the grain shape during sintering completely tetrakaidecahedron, not before late sintering stage is reached. More importantly for the BCC the pore unit shape is octahedral, while for the HCP and FCC packings the pores are tetrahedral and octahedral respectively, which is different from the assumed cylinder and spherical pore geometry assumed by Coble.

The densification behavior was also simulated for a simulation temperatures of 1:1:13 and attempt frequencies of 1:1:1 for the grain growth, pore migration and vacancy formation temperatures, respectively. These values have previously been used to model the sintering of copper spheres \cite{Tikare_2010}. For a random packing powder the density at which grain growth occurred decreased to $\rho=0.83$ for this temperature set. No change in the sintering behavior or the microstructural evolution was seen for the BCC, FCC and HCP samples, except that the time required to reach e.g. a relative density of $0.995$ was 2.0,  2.1 and 1.3 times longer for the BCC, FCC and HCP samples, respectively. This invariance to temperature was also observed experimentally by Barringer \& Bowen \cite{Barringer_1982} for the mono-sized uniformally packed TiO$_2$.

In conclusion the sintering behavior of close packed spheres was investigated for body centered cubic (BCC), face centered cubic (FCC) and hexagonal closed packed spheres (HCP). All systems showed no grain growth until full density was reached. This is a very significant ability, caused by the regular packing of the powder. All grains evolved to a tetrakaidecahedron shape with coordination number 14 at full density, similar to the geometry considered by Coble. The dominant closed pore type in BCC was the pore with coordination number four and not six as expected, while for the FCC and HCP packings the dominating closed pore type had coordination number of six. These systems thus shows the great importance of the packing of the initial powder compact and demonstrates the potential improvement in sintering behavior gained by regular packing of the powder. These systems are also ideal candidates for standard modeling systems in sintering.

The authors would like to thank the Danish Council for Independent Research Technology and Production Sciences (FTP) which is part of The Danish Agency for Science, Technology and Innovation (FI) (Project \# 09-072888) for sponsoring the OPTIMAC research work.  Sandia National Laboratories is a multi-program laboratory managed and operated by Sandia Corporation, a wholly owned subsidiary of Lockheed Martin Corporation, for the U.S. Department of Energy's National Nuclear Security Administration under contract DE-AC04-94AL85000.

\end{document}